\documentclass[a4paper,11pt]{article}
\usepackage{jinstpub} 
\usepackage{lineno}
\usepackage{caption}
\usepackage{subfig}
\usepackage{multirow}
\usepackage{siunitx}
\usepackage{hyperref}
\usepackage[dvipsnames]{xcolor}
\usepackage{soul}

\newcommand{\red}[1]{\color{black}{#1}}

\title{AEgIS Experiment at CERN: Design and Commissioning of SARA (Scintillator Assemblies to Reveal Annihilations)}

\author[a,b,1]{P. Conte\note{Corresponding author.}}
\author[a,b]{G.~Consolati}
\author[c]{M.~Prata}
\author[d]{M.~Berghold}
\author[e]{R.~Caravita}
\author[e,f]{R.~Ferguson}
\author[g]{M.~Grosbart}
\author[d,e,f]{F.~Guatieri}
\author[g]{S.~Haider}
\author[g]{G.~Khatri}
\author[h]{L.~Lappo}   
\author[i,j]{P.~Moskal} 
\author[d]{M.~M\"{u}nster}
\author[e,f]{L.~Penasa}
\author[i,j]{S.~Sharma}

\affiliation[a]{INFN Milano,\\ via Celoria 16, 20133~Milano, Italy}
\affiliation[b]{Department of Aerospace Science and Technology, Politecnico di Milano,\\ via La Masa 34, 20156~Milano, Italy}
\affiliation[c]{INFN Pavia,\\ via Bassi 6, 27100~Pavia, Italy}
\affiliation[d]{Heinz Maier Leibnitz Zentrum (MLZ), Technical University of Munich,\\ Lichtenbergstraße 1, 85748, Garching, Germany}
\affiliation[e]{TIFPA/INFN Trento,\\ via Sommarive 14, 38123~Povo, Trento, Italy}
\affiliation[f]{Department of Physics, University of Trento,\\ via Sommarive 14, 38123~Povo, Trento, Italy}
\affiliation[g]{Physics Department, CERN,\\ 1211~Geneva~23, Switzerland}
\affiliation[h]{Warsaw University of Technology, Faculty of Physics,\\ ul. Koszykowa 75, 00-662, Warsaw, Poland}
\affiliation[i]{Marian Smoluchowski Institute of Physics, Jagiellonian University,\\  ul. \L ojasiewicza 11, 30-348  Krak\'ow, Poland} 
\affiliation[j]{Centre for Theranostics, Jagiellonian University,\\ ul. Kopernika 40, 31-501 Krak\'ow, Poland}

\emailAdd{pietro.conte@mi.infn.it}

\abstract{
SARA is the system of plastic scintillators coupled with silicon photomultipliers that will take part in the AEgIS experiment at CERN, measuring the time-of-flight of antihydrogen as it falls through a moiré deflectometer. Its development focused on simplicity, versatility and economy of the design and was supported by both physical tests and numerical simulations. The instrument structure pairs the utilization of the scintillators as structural components with custom made 3D printed corner elements and the electronics allows {\red selection} between coincidence discrimination made on each scintillator and made between different scintillators.
}


\begin{document}
\maketitle
\flushbottom

\section{Introduction}
\label{intro}

The AEgIS experiment (Antihydrogen Experiment: gravity, Interferometry, Spectroscopy) at CERN aims to measure the effects of Earth’s gravity on antimatter in the absence of magnetic field influences{\red ,} with an accuracy of 1\% \cite{Objective}. The results will pave the way for testing the Weak Equivalence Principle of General {\red R}elativity and the Charge, Parity and Time Reversal Symmetry embedded in Quantum Field Theory \cite{AEgIS}.

\begin{figure}[h]
    \centering
    \includegraphics[width=0.93\linewidth]{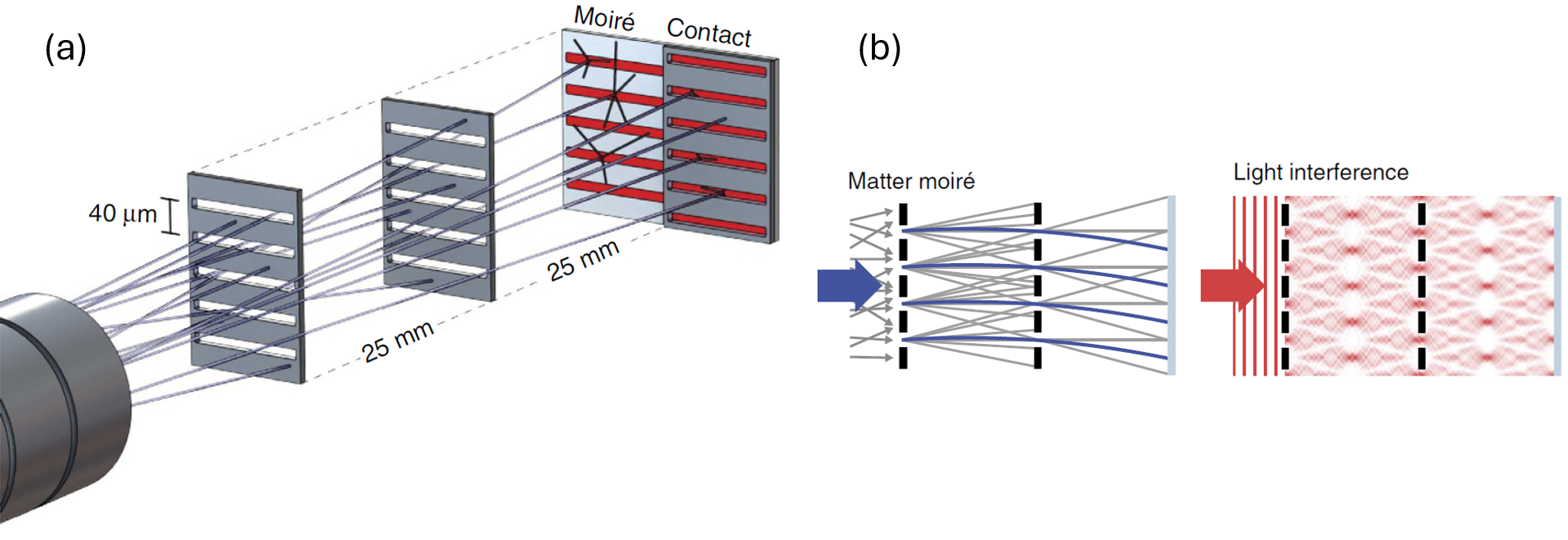}
    \caption{{\red (a) (b) Deflectometer functioning principle, adapted from \cite{defl_pic}, where antihydrogen atoms pass through two gratings and form a fringe pattern.} }
    \label{fig_aegis01}
\end{figure}

{\red The experiment will use a horizontally-boosted, free-falling pulsed beam of antihydrogen ($\bar H$), which will be sent through a moiré deflectometer contained within a horizontal tube maintained under ultra-high-vacuum conditions. Two horizontal gratings, whose relative positions can be tuned, are placed at the entrance and at the midpoint of the tube, perpendicular to its axis. They generate a periodic spatial distribution of antihydrogen at a fixed distance downstream of the second grid, since only antiparticles following nearly horizontal trajectories avoid colliding and annihilating with the gratings } (Figures \ref{fig_aegis01} (a) and (b)).

{\red This apparatus differs from a Mach-Zehnder atom interferometer, that operates in a regime where the de Broglie wavelength $\lambda$ is much larger than the grating period $d$, and where three identical gratings are placed at an equal distance $L$ from each other. On the third one, an interference pattern is produced, showing the same period as the gratings \cite{Mach1} \cite{Mach2}. In the presence of gravity, this pattern is shifted vertically by a distance equal to:
\begin{equation}
    \Delta y = -g\Delta t^2
    \label{eq1}
\end{equation}

where $g$ is the local gravitational acceleration and $\Delta t = v/L$ is the time-of-flight of a particle beam moving at velocity $v$ between two gratings. The use of a Mach-Zehnder interferometer is problematic in the presence of antimatter due to various decoherence effects. More important, the divergence of the antiatom beam must be less than the diffraction angle $\alpha = \lambda/d$. A way to circumvent this limitation is indeed to operate the device in the moiré regime, where $\alpha \ll d/L$, and consequently the diffractive effects are replaced by a classical shadow pattern of the antiatoms. Once past the two gratings, the latter converges on a final detector (either a third grating and a counter, as for the Oberthaler original configuration \cite{Mach3}, or a position-sensitive detector, as for the AEgIS experiment). Nevertheless, the law of motion of the free-falling atoms (Eq. \ref{eq1}) is valid in both regimes. 

In the present setup,} the fringe pattern -- {\red a fringes-like distribution of annihilation points} -- obtained after the second grating will be collected by OPHANIM, a high-resolution{\red ,} position-sensitive{\red ,} 3840-Megapixel CMOS detector \cite{OPHANIM} capable of comparing the antihydrogen fringe pattern with {\red the one} of a reference light source{\red , which has} traveled the same path and {\red has been negligibly affected by gravity.} 

The vertical position difference measured by the detector will be {\red used} into the equation $\Delta y = -\bar{g}\Delta t^2$ (derived from eq. \ref{eq1}), giving as result the intensity of the gravity acceleration acting on antimatter $\bar g$. The antihydrogen transit time will be measured as the difference between the time instant {\red of its arrival} at OPHANIM and the one {\red of its formation}, known with an uncertainty of \SI{250}{\nano\second} \cite{formation}. Antihydrogen will be obtained by {\red exploiting} the interaction between antiprotons, coming from the Antimatter Decelerator (AD), and excited positronium in Rydberg states ($Ps^*$), generated by shooting two laser beams (\SI{205.045}{\nano\meter} and \SI{1693}{\nano\meter}) toward a cloud of ortho-positronium. The latter will be formed through the interaction between a positron beam, coming from a $^{22}Na$ source, and a {\red nanochanneled silicon} target \cite{formation2}. This antihydrogen production is called charge-exchange process and {\red can be} summarized by the equation \cite{formula}:
\begin{equation}
    Ps^* + \bar{p} \rightarrow \bar{H}^* + e^- 
\end{equation}


\begin{figure}[h]
    \centering
    \includegraphics[width=1\linewidth]{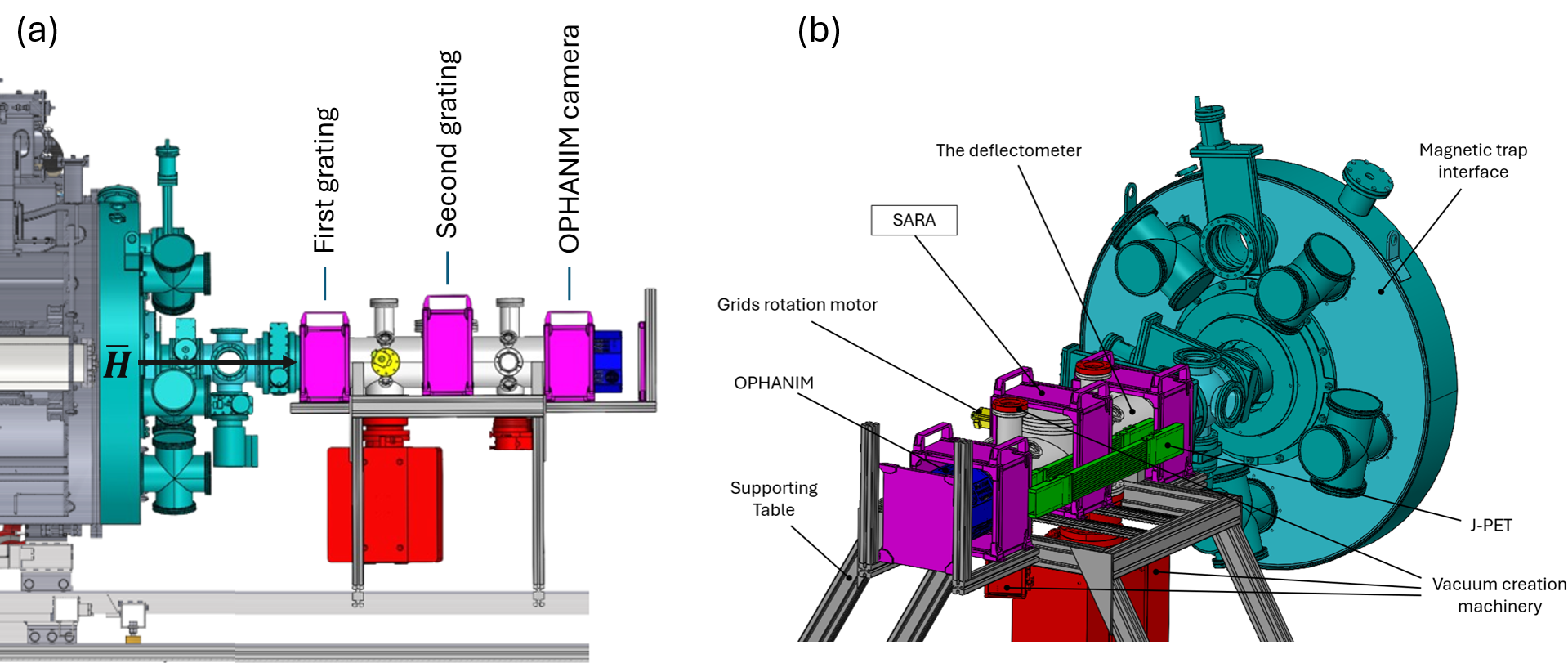}
    \caption{{\red (a) Gravity Module of the AEgIS apparatus, where the pink boxes indicate the scintillator assemblies (SARA modules) positioned around the gratings and OPHANIM. (b) The complete Gravity Module layout, in which the pink boxes represent scintillator assemblies and the green slabs correspond to the planned J-PET modules \cite{JPET} }}
    
    \label{fig_aegis02}
\end{figure}

{\red This paper presents and discusses} SARA (Scintillator Assemblies to Reveal Annihilations), the system of scintillators that aims to measure the time-of-flight distribution of antihydrogen atoms as they move through the moiré deflectometer {\red of the AEgIS apparatus (Figure \ref{fig_aegis02} (a) and (b))}. {\red Timing measurements} {\red are} essential {\red for determining} the gravitational acceleration of antimatter and {\red will be performed} by detecting the annihilation products generated {\red by} the interaction between {\red a fraction of} the antiatoms {\red with} the two gratings and {\red with} OPHANIM.

\section{Objective and Requirements}

\label{reqs}

{\red As anticipated,} the main goal of the SARA detector is to measure the time-of-flight $\Delta t$ distribution of antihydrogen atoms by detecting their annihilation products {\red (}in particular pions{\red )} generated when they collide with either one of the two moiré gratings or the OPHANIM {\red detector} surface. {\red This data will allow working out $g$ from Eq. \ref{eq1} in combination with the vertical deflection $\Delta y$ measurement from the OPHANIM detector. In addition, the number of annihilations produced on the two grids and recorded by SARA will allow to estimate the fraction of $\bar H$ produced that do not reach the final CMOS detector, providing information on the temporal spread of the antihydrogen beam and the transmittivity of the individual gratings. 
The 1\% accuracy goal depends on both the minimal velocity of the antihydrogen beam and the accumulated statistics (i.e., the total number of antiatoms reaching the final detector, and therefore on the production efficiency), making it crucial for the SARA detector to detect antihydrogen annihilations with a high efficiency.}

{\red The SARA detector was henceforth designed following four paradigms}:
\begin{enumerate}
    \item {\red The detector shall clearly distinguish the annihilations occurring at the two gratings and at OPHANIM separately, so it must be composed of at least three independent modules.}
    \item {\red Each module must be able to detect the majority of annihilation events happening at the assigned annihilation site. Therefore, each module shall detect the majority of the pions produced during the nucleon-antinucleon annihilation, so their efficiencies have to be higher than $50\%$, to be in line with the 58\% efficiency of the OPHANIM detector and not impose more stringent requirements on the antihydrogen source flux.}
    \item {\red The time resolution of each detecting element of the modules should be on the order of \SI{10}{\nano\second}, to allow a measurement of the $\bar H$ time-of-flight at the same order-of-magnitude accuracy as their formation instant, defined by laser pulses duration (4-5 \SI{}{\nano\second}), and the spread in time-of-flight of the $Ps^*$ atoms in the antiproton cloud (10-50 \SI{}{\nano\second}). }
    \item {\red The SARA detector should allow rejecting both dark counts, cosmic-ray muons and natural radioactivity of the laboratory by allowing to reveal coincidences, ensuring that the events originated in its encompassed volume are easily distinguishable.}
    
\end{enumerate}

\section{Detector Design}
\label{design_sec}

\subsection{Initial Design Assumptions}

Starting from the requirements {\red outlined above}, {\red a} combination of plastic scintillators and Silicon Photomultipliers (SiPMs) {\red was adopted, since it is an effective technological choice to reveal annihilations}. 
The former {\red were chosen} both for their fast rise time (often {\red below} \SI{1}{\nano\second}), their high detection efficiency {\red and their reduced costs in encompassing large volumes}. {\red According to the results of the ATRAP Experiment \cite{atrap},} a thickness of around \SI{1}{\centi\meter} {\red generates} a sufficient amount of photons during the transit of pions {\red produced by annihilation}.
On the other hand, the SiPMs were selected after a trade off {\red against} photomultiplier tubes (PMTs) as alternative. Even {\red though} PMTs offer superior performance in terms of dark count rates at room temperature, SiPMs are notoriously {\red insensitive to} magnetic fields \cite{SiPM_Mag}{\red , observable in the surrounding of SARA due to the nearby presence of the \SI{1}{\tesla} and \SI{5}{\tesla} magnets in the main AEgIS cryostat. Indeed, in the nearby region magnetic fields of around \SI{1}{\milli\tesla} were measured, which can therefore influence the gain of a PMT while having no effect on SiPMs. The deflectometer will, however, be protected by suitable mumetal shields to avoid affecting the trajectories of the antihydrogen atoms.} {\red This} means that, unlike PMTs, {\red SiPMs} do not require complex shielding, keeping the detector footprint compact; in addition, PMTs usually require {\red cumbersome} light guides {\red which would need more space than that actually available between adiacent scintillators (around \SI{25}{\centi\meter}), unless one resorts to rather complex geometries.}

\begin{figure}[h]
    \centering
    \includegraphics[width=0.97\linewidth]{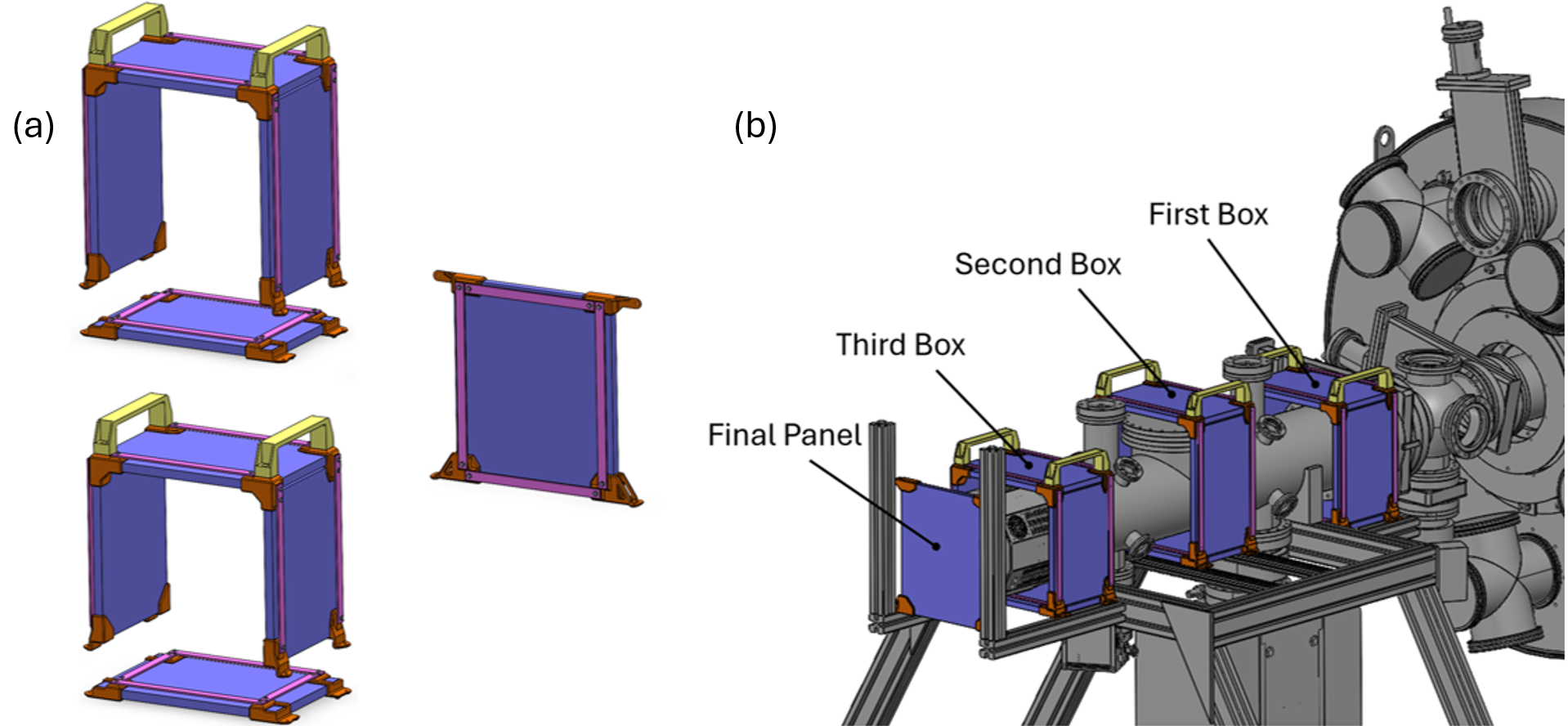}
    \caption{ (a) Big box and one small box showing the two parts architecture; on the right the final panel, (b)  SARA integrated in the AEgIS apparatus.}
    \label{structure01}
\end{figure}

In order to maximize the detection efficiency and {\red effectively minimize} the atmospheric muons background noise, while keeping the design as simple as possible, twelve rectangular plastic scintillators {\red were assembled} in three boxes to be placed around the gratings and the OPHANIM {\red detection} surface, partially covering the solid angle around them{\red .} {\red An additional} rectangular scintillator panel {\red was placed} behind the final detector to enhance its solid angle coverage; the {\red final assembly and placement are shown} in Figure \ref{structure01} (b). 
The {\red need} of performing baking procedures {\red in order} to obtain ultra{\red -}high vacuum inside the tube, coupled with the high sensitivity of {\red the selected scintillator material} to high temperatures {\red (see Subsection \ref{scints})}, imposed the definition of a box architecture that allows fast integration and removal procedures. A test was performed to {\red determine} whether the application of mica insulation layers between the deflectometer and the scintillators could {\red prevent} material damages{\red .} {\red However,} it was found that it is impossible to leave SARA intact during the baking. 
{\red Consequently, the chosen boxes architecture consists of two parts:} an upper {\red section}, containing the two vertical and the upper horizontal scintillators, and a lower {\red section}, containing the remaining panel (Figure \ref{structure01} (a)). {\red The choice to divide each box in two sections allows a fast removal (and subsequent mounting) when a baking procedure for the whole deflectometer is needed. This ensure that this procedure will not be a bottleneck of the experiment execution.} To integrate {\red the} boxes on the instrument, first the lower part {\red is} placed in its correct position and then the upper part {\red is} lowered from above{\red .} Finally, the two parts are fastened both together and to the instrument supporting structure {\red by applying four screws at the four bottom corners of the created box.}

The whole deflectometer apparatus, {\red including} SARA, will lay on a single supporting table made of Bosch profiles with various section dimensions. In particular, each box of scintillators, after the integration in the moiré {\red deflectometer}, will be supported by two horizontal and parallel beams with section variable between $20$x\SI{20}{\milli\meter\squared}, $30$x\SI{30}{\milli\meter\squared} and $45$x\SI{45}{\milli\meter\squared} sizes{\red .} {\red On the other hand,} the final panel will be supported by one or two horizontal beams {\red and} two vertical beams for stability{\red , therefore} SARA {\red has been} designed {\red to guarantee} adaptability to various {\red Bosch profiles} section dimensions and to {\red supporting} beams orientation parallel or orthogonal to the deflectometer axis.

\subsection{Scintillators Properties}
\label{scints}

The material selected to create the scintillators is the general purpose BC-404 from Luxium Solutions, of which some \SI{1}{\meter} long, \SI{277}{\milli\meter} wide, \SI{13}{\milli\meter} thick panels were already available and whose properties are collected in Table \ref{ATRAP}.

\begin{table}[h]
    \caption{{\red Properties of } BC-404 scintillator \cite{scint1}, \cite{scint2}.}
    \centering
    \begin{tabular}{rl}
    \hline
        Property & Value  \\
        \hline
         Name: & Luxium Solutions BC-404 \\
         Base: & Polyvinyl-toluene \\
         Panels dimensions (\SI{}{\milli\meter}): &  $1000$x$277$x$13.3$ \\
         Refractive index: & $1.58$ \\
         {\red Softening point (\SI{}{\celsius}):} &  {\red 70}  \\
         Rise time (\SI{}{\nano\second}): & $0.7$ \\
         Decay time (\SI{}{\nano\second}): & $1.8$ \\
         Wavelength of max. emission (\SI{}{\nano\meter}): & $408$ \\
         Light attenuation length (\SI{}{\centi\meter}): & $140$ \\
         Light output (n. photons/MeV): & $10800$\\
         \hline
    \end{tabular}
    \label{ATRAP}
\end{table}

\begin{figure}[h]
    \centering
    \includegraphics[width=0.95\linewidth]{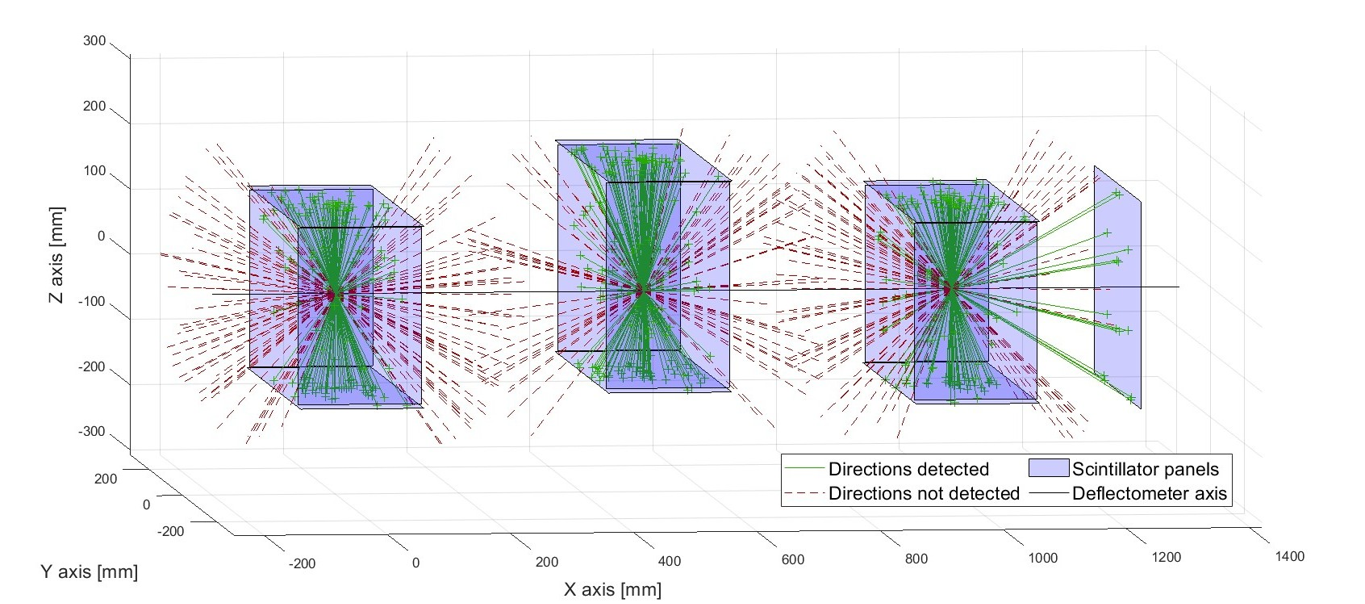}
    \caption{{\red Monte Carlo simulation used to determine the solid angle fraction of each SARA module. The green lines are the directions revealable by the detector modules while the red dotted lines are directions not revealed by the module placed around their generation point.} }
    \label{solid ang}
\end{figure}

Given the limited space around the deflectometer, the on-axis dimension of the {\red boxes} scintillator panels was fixed to \SI{200}{\milli\meter} {\red and their last dimension was determined according to the detector system surrounding. Moving to the thirteenth panel (placed behind OPHANIM), its dimensions were derived from the starting BC-404 material panels plus the decision of having this scintillator centered on the deflectometer axis. Concerning the sizes, all the horizontal scintillators of the boxes are \SI{316}{\milli\meter} long, the second box has vertical scintillators \SI{316}{\milli\meter} high and the first and third boxes have vertical scintillator panels \SI{270}{\milli\meter} high. On the other hand, the final scintillator is \SI{275}{\milli\meter} wide and \SI{318}{\milli\meter} high. A simple Monte Carlo simulation was exploited to estimate the fraction of the solid angles around each annihilation region that SARA is able to cover. In particular, it considered $10^6$ randomly generated directions per annihilation site and approximated these lasts as the points of intersection between the gratings or OPHANIM and the deflectometer axis. In addition, the analysis computed the cross detection (e.g., annihilation products generated at the first grating and detected by the second box), finding it is negligible. The results are reported in Table \ref{solid} and Figure \ref{solid ang}.}

{\red The solid angles coverage computed, coupled with the second requirement presented in Section \ref{reqs}, led to the finding of a minimum acceptable value for the scintillator panel efficiencies. In particular, for the first box it was found a minimum averaged efficiency of 0.755, for the second a minimum averaged value of 0.780 and for the third box plus final panel assembly it was found a minimum averaged value of 0.703.}

\begin{table}[h]
    \centering
        \caption{{\red Solid angle fractions around the three annihilation sites that are covered by the three modules}.}
    \begin{tabular}{ccc}
    \hline
        Annihilation {\red site} & Module considered & Solid angle fraction (\%) \\
        \hline
        First grating & First box & $66.2$ \\
        Second grating & Second box & $64.1$ \\
        OPHANIM & Third box and final panel & $71.1$\\
        \hline
        {\red First grating} & {\red Second Box} & $1.41$\\
        {\red First grating} & {\red Third box and final panel} &  $0.96$\\
        {\red Second grating} & {\red First box} &  $1.21$ \\
        {\red Second grating} & {\red Third box and final panel} &  $1.97$\\
        {\red OPHANIM} & {\red First box} &  $0.16$ \\
        {\red OPHANIM} & {\red Second box} &  $1.37$\\
        \hline
        
    \end{tabular}
    \label{solid}
\end{table}

\subsection{Detector Mechanical Design}

Given the thickness of the scintillator panels, they have been exploited as structural elements for the boxes, obtained by connecting the panels by means of custom 3D printed corner elements, by some aluminum bars and two handles.
The material selected to 3D print the corner elements is Nylon PA12 \cite{nylon}, chosen for its good mechanical properties over economic cost ratio, and the printing technology selected was the Multi Jet Fusion, that guarantees good isotropy of the final products. 
A trade-off analysis was conducted to compare a plastic 3D printed corner elements solution with a design obtained by welding metal plates. {\red Despite it would have been more economic}, the latter solution was discarded because of the long-term damages at the points of contact between the scintillators and the metal plates edges. {\red In addition,} the first alternative had also the advantage of minimizing the overall {\red detector} weight, important for its supporting structure architecture.

\begin{figure}[h]
    \centering
    \includegraphics[width=0.94\linewidth]{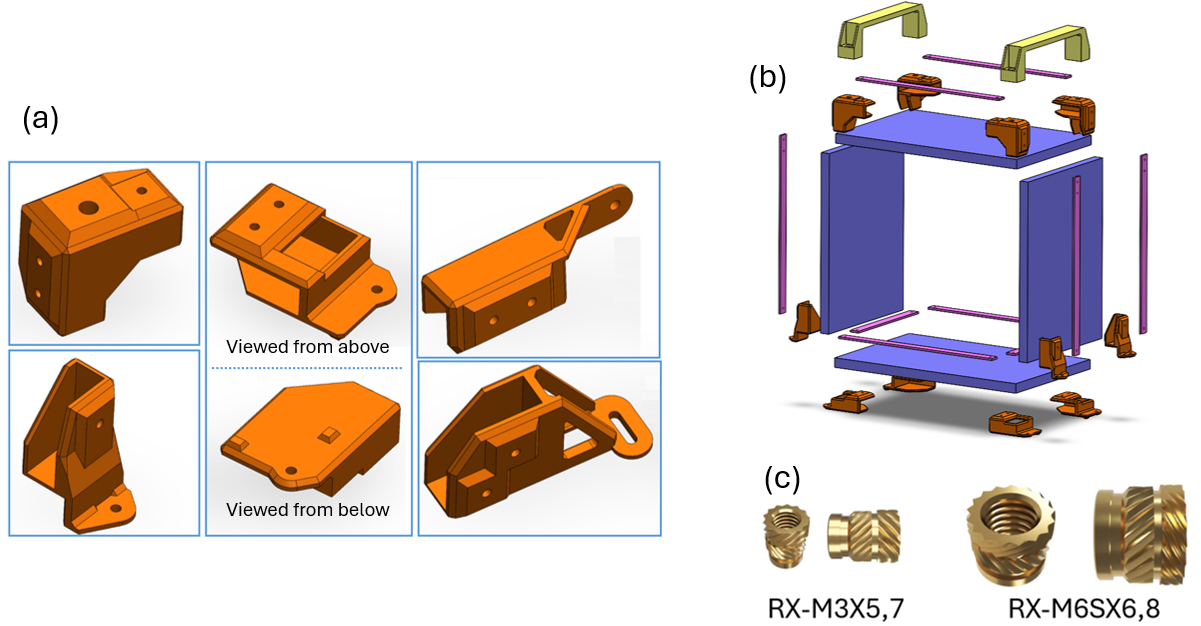}
    \caption{(a) The variants of corner elements{\red , coming in two mirrored versions, as can be seen in the central image} (b) Exploded view of the first box, (c) The {\red threaded inserts} selected.}
    \label{structure02}
\end{figure}

The shape of the corner elements was determined according to their structural function, their role in protecting the scintillators corners from accidental impacts, the minimization of their mass and the simplicity of the 3D printing process. {\red Furthermore}, the design process focused on the realization of as {\red many identical} components as possible, feature that allowed to reduce their unitary cost. The result is shown in Figure \ref{structure02} (a).

Moving to the aluminum bars and handles, which hold the corner elements in place and facilitate handling, they have both been selected among COTS products. Aluminum $6060$ T5, $15$x\SI{3}{\milli\meter} section, \SI{2}{\meter} long bars have been considered{\red ;} for the latter{\red ,} the Elesa M.443/200-CH-C9 handles were chosen. To connect corner elements, aluminum bars and handles, the {\red former} have been equipped with ensats, brass cylindrical inserts with a thread on the inside and designed to be hot-inserted in dedicated holes inside plastic components{\red .} {\red T}he only disadvantage of this solution is to impose a minimum depth and walls thickness for the insertion holes, which slightly conflicts with the minimization of the printed material. In Figure \ref{structure02} (c), the two selected products from Ruthex are shown. 

\subsection{Finite Elements Mechanical Simulations}
\label{FEM_sec}

{\red In order to test the structural components before their physical realization,} the whole design process {\red was supported} with mechanical finite elements analysis, using Autodesk Fusion as simulating software. In particular, the focus was directed to the central box, which is the bulkiest and so the one representing the critical case for the mechanical stresses. The analysis investigated four loading conditions {\red characterized by the combination of weight (acting on every point of the structure) and a horizontal force (applied in a certain point of the structure),} {\red simulating} both the nominal and the overloaded scenarios described in Table \ref{FEM scenarios}.

\begin{table}[h]
    \centering
        \caption{The two scenarios considered in the FEM analysis.}
    \begin{tabular}{rcc}
    \hline
        Quantity & Nominal scenario & Overloaded scenario \\
        \hline
         {\red Weight W (\SI{}{\newton\per\kilogram})} & $9.81$ & $13.24$ \\
         {\red Horizontal force F (\SI{}{\newton})} & $10.0$ & $15.0$ \\
         \hline
    \end{tabular}
    \label{FEM scenarios}
\end{table}

\begin{figure}[h]
    \centering
    \includegraphics[width=0.9\linewidth]{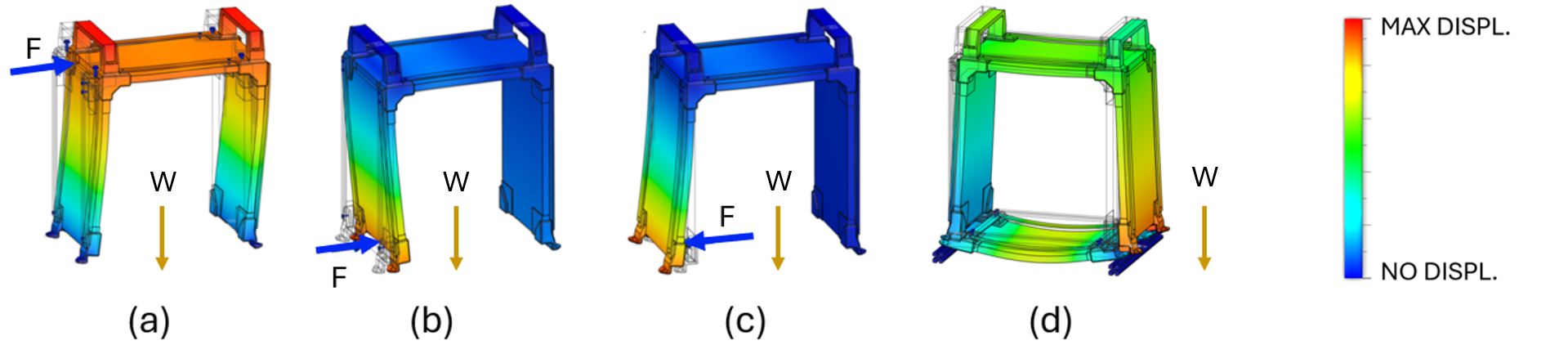}
    \caption{FEM analysis computed deformations. The blue arrows represent the applied force and gravity is {\red always} directed downward (the displacement is exaggerated for illustration purposes).}
    \label{fem}
\end{figure}

In the first loading condition, {\red the box,} already fastened to the instrument supporting structure, is loaded with a horizontal force applied on one of its upper {\red shorter horizontal} edges. This load condition can also occur when only the upper part is fastened to the box supports and it simulates an accidental impact with the box after it has been integrated into the instrument. 
In the second and third cases, a horizontal load is applied respectively inward and outward at a lower edge of the upper part. These two load combinations simulate an accidental impact that can happen during the upper part integration or removal. 
Finally, in the fourth load condition the central box is supported by two $20$x\SI{20}{\milli\meter\squared} {\red section} Bosch profiles that run parallel to the deflectometer axis. 
In this case the gap between the two supporting beams is {\red larger} than {\red the length of the lower horizontal scintillator, meaning that} the whole box is {\red sustained} by the {\red four horizontal flanges of the lower part} corner elements {\red only (the single holed flange appreciable in the central image of Figure \ref{structure02} (a)).
 This last load combination does not take any horizontal force into account.}
Given the high {\red elongation} at break of Nylon PA12 (20\%), maximum displacement was used as an indicator of stiffness. {\red Indeed, the safety factor computed, evaluated as $SF = \sigma_Y/\sigma_W$, where $\sigma$ indicates stresses, Y means Yield and W stands for Work, was always above 6.} The results of the analysis{\red ,} reported in Figure \ref{fem} and Table \ref{deformations_results}{\red , show that the deformations can be considered reasonably acceptable.}

\begin{table}[h]
    \centering
       \caption{Maximum displacements computed with the FEM analysis.}
    \begin{tabular}{ccc}
    \hline
        Case & Nominal Scenario (\SI{}{\milli\meter}) & Overloaded Scenario (\SI{}{\milli\meter}) \\
        \hline
         (a) & $0.443$ & $0.670$ \\
         (b) & $0.997$ & $1.497$ \\
         (c) & $1.659$ & $2.497$ \\
         (d) & $0.122$ & $0.133$\\
         \hline
    \end{tabular}
    \label{deformations_results}
\end{table}


\subsection{Readout Electronics}
\label{Electronics_sec}

\begin{figure}[h]
    \centering
    \includegraphics[width=0.87\linewidth]{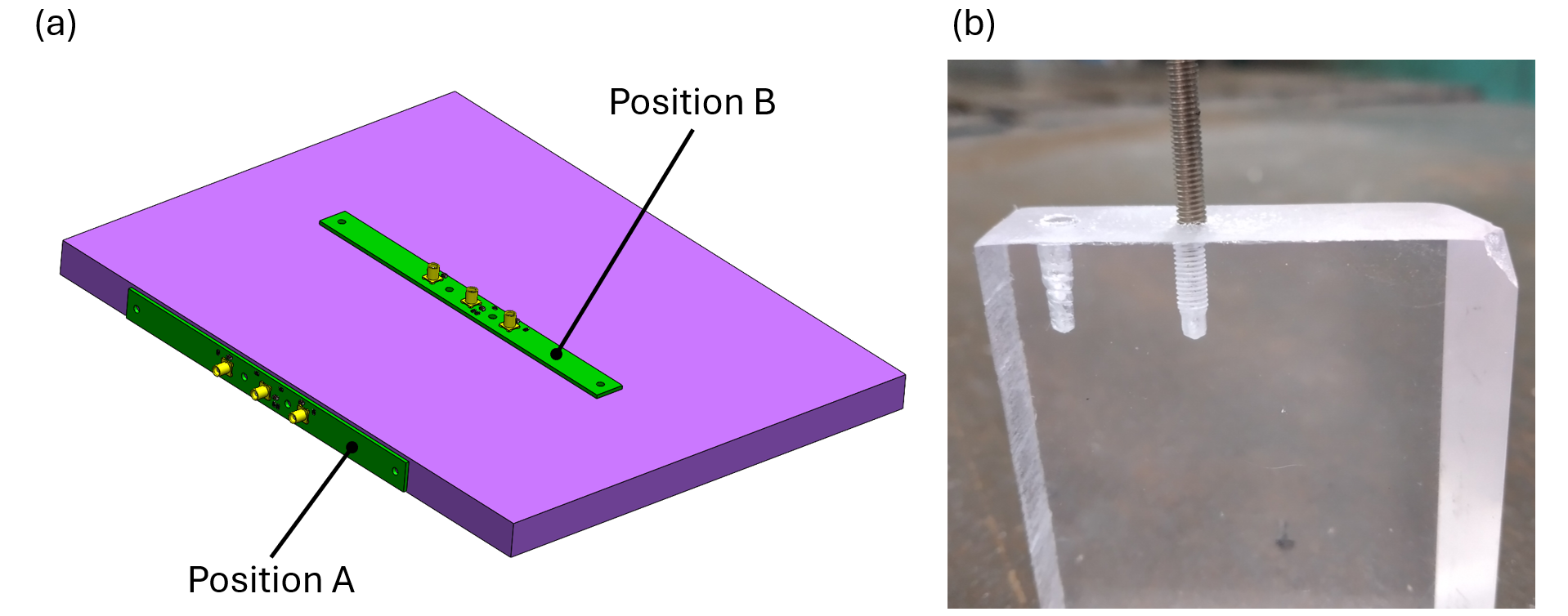}
    \caption{(a) The two PCB positions tested, (b) The result of the threaded hole creation test.}
    \label{position hole}
\end{figure}

Concerning the electronics design, on every scintillator panel one PCB hosting twelve SiPMs {\red was applied. The SiPMs were} divided into two groups of six, each one equipped with a dedicated {\red signal} output port, with the possibility to read them either in coincidence or in parallel. It was initially decided to mount each PCB on one lateral side of the rectangular scintillators, using nylon screws as fastening strategy due to {\red its} reversibility, effectiveness and low cost. For this reason, the PCBs have been designed to be elongated, as wide as the thickness of the scintillators and equipped with four screw holes. 
A preliminary test was performed {\red to compare} the average output amplitude of a PCB hosting 3 SiPMs first mounted on a lateral side of a SARA scintillator and then on one of its two main surfaces, as represented in Figure \ref{position hole} (a). The high-energy particles source used was atmospheric muons and the result of the two acquisitions was practically identical. Consequently, the lateral mounting position was selected for the better cable management granted and {\red based} on the reasonable assumption that a SiPM {\red applied on} a lateral surface has a direct view on a wider volume of the scintillator {\red with respect} to one {\red applied to} a main surface.
{\red Then}, a test was carried out to determine how to create threaded holes in the scintillator panels; the result is shown in Figure \ref{position hole} (b). It is possible to {\red generate} the holes, but the procedure developed is time-consuming and {\red quite} complex{\red ;} moreover{\red ,} any error could damage the entire scintillator panel due to its extreme sensitivity to heat{\red . F}or this reasons{\red ,} it was concluded to {\red use} black {\red adhesive} tape as a still reversible and {\red straightforward} solution to secure the PCBs {\red and to add an additional layer of stray light protection to the SiPMs. In particular, compressing tape pieces were applied to press the two SiPMs regions of the PCB onto the scintillator, and covering pieces closed the gap between the circuit board and the scintillator thus preventing light infiltrations; also the PCB screw holes were taped as well.}

The {\red selected} SiPM is the Hamamatsu S14160-3050HS \cite{SiPM}. The choice was mainly driven by its peak sensitive wavelength, which closely matches the maximum emission wavelength of BC-404, and its compact dimensions ($3.4$x$3.4$x\SI{1.35}{\cubic\milli\meter}), tailored for the lateral side application. Some of the SiPM properties are collected in Table \ref{SiPM tab}.

\begin{table}[h]
    \centering
        \caption{Characteristics of the Hamamatsu {\red SiPM} S14160-3050HS}
    \begin{tabular}{rcc}
    \hline
        Quantity & Value & Unit of measure \\
        \hline
         Effective photosensitive area & $3.0$x$3.0$ & \unit{\square\milli\meter} \\
         Window refractive index & $1.57$ & - - \\
         Spectral response range  & $270$ to $900$ & \unit{\nano\meter} \\
         Peak sensitive wavelength & $450$ & \unit{\nano\meter} \\
         Recommended operational voltage & $38$ + $2.7$ & \unit{\volt} \\
         Dark current (nominal - maximum) & $0.6$ - $1.8$ & \unit{\micro\ampere} \\
         Gain & $2.5$ x $10^6$ & - - \\
         Terminal capacitance & $500$ & \unit{\pico\farad} \\
                                   \hline
                                   
    \end{tabular}

    \label{SiPM tab}
\end{table}

\begin{figure}[h]
    \centering
    \includegraphics[width=0.95\linewidth]{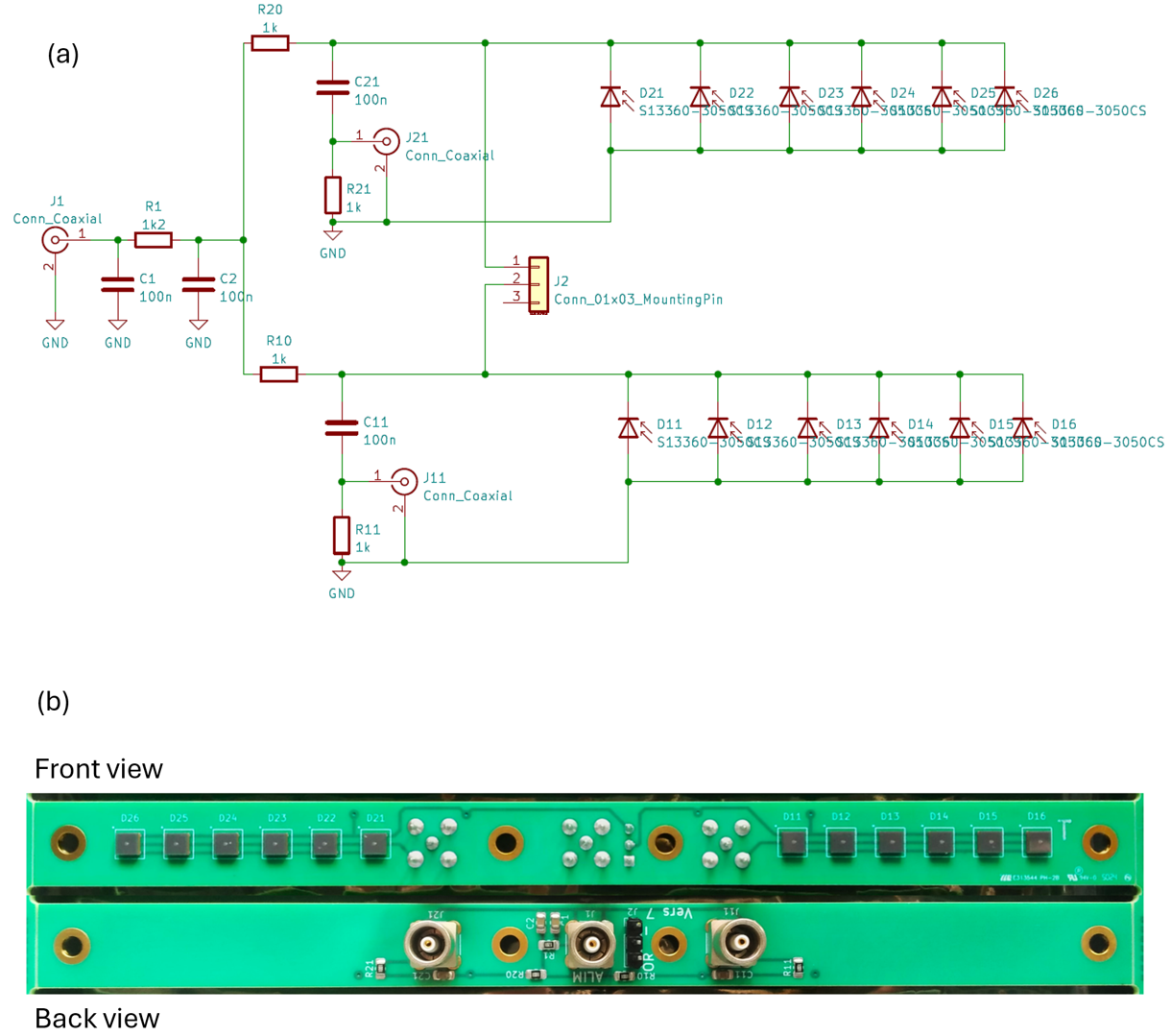}
    \caption{(a) PCB scheme (b) Front and back views of a manufactured PCB}
    \label{pcb}
\end{figure}

The PCB architecture, shown in Figure \ref{pcb} (a), is based on two channels, each consisting of six parallel SiPMs powered by a single supply line (J1 in the schematic shown in Figure \ref{pcb} (a)) operating at 4 V above the breakdown voltage. After the power-line filtering stage, each channel includes a branch with a $100$ \unit{\nano\farad} discharging capacitor (C11 and C21 in Figure \ref{pcb} (a)), which enables measurements of a single output signal as the voltage variation across a 1 \unit{\kilo\ohm} series resistor (R11 and R21 in Figure \ref{pcb} (a)). The goal for the PCB architecture design was to obtain a straightforward solution. 
In order to enable the 12 SiPMs in-parallel reading strategy, a short-circuit bridge can be {\red obtained by applying a jumper to a central 3 pins connector (J2 in the figure)}{\red . M}oreover, due to the particular shape selected for the PCBs, the SiPMs have been {\red arranged} to form {\red a pair of} six elements strips {\red aligned with} the PCB main dimension (Figure \ref{pcb} (b)). To improve the interaction between the detectors and the plastic scintillators, an optical grease {\red (the EJ-550 from Eljen Technology \cite{OptGrease}) was applied} at {\red their} contact surface{\red . Indeed,} a test confirmed {\red that, in the presence of grease and during the same amount of time,} an increase in the number of detected scintillations {\red caused by muons is observed}.

\begin{figure}[h]
    \centering
    \includegraphics[width=1\linewidth]{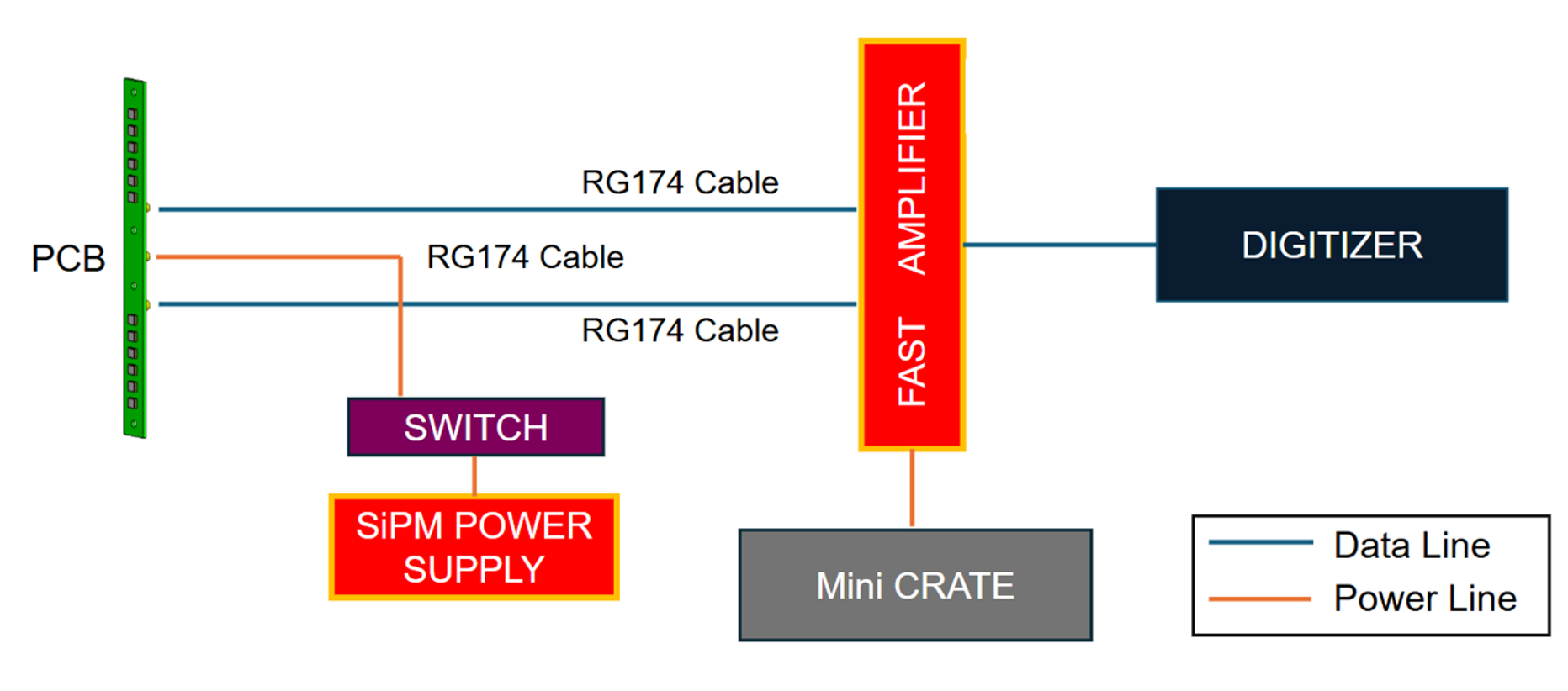}
    \caption{Scheme of the electronics architecture. }
    \label{electronics arc}
\end{figure}

{\red Concerning} the overall electronics setup, all the PCBs {\red are powered} {\red by} a single SiPMs power supply, the CAEN DT5485PB \cite{Powersup}, exploiting a single input-$13$ outputs custom-made switch. {\red A}ll the PCBs outputs {\red are} collected by fast amplifiers {\red with gain equal to 10}, the CAEN N979 \cite{Amplifier}, that subsequently send the amplified signals to {\red two or four} $14$-bit, {\red 250 MS/s} digitizers, {\red the CAEN 5725SB} \cite{Digitizer}, for data acquisition. Both the power and data lines use coaxial cables (RG174) with LEMO connectors. A representation of the electronics architecture is reported in Figure \ref{electronics arc}. 

\begin{figure}[h]
    \centering
    \includegraphics[width=0.95\linewidth]{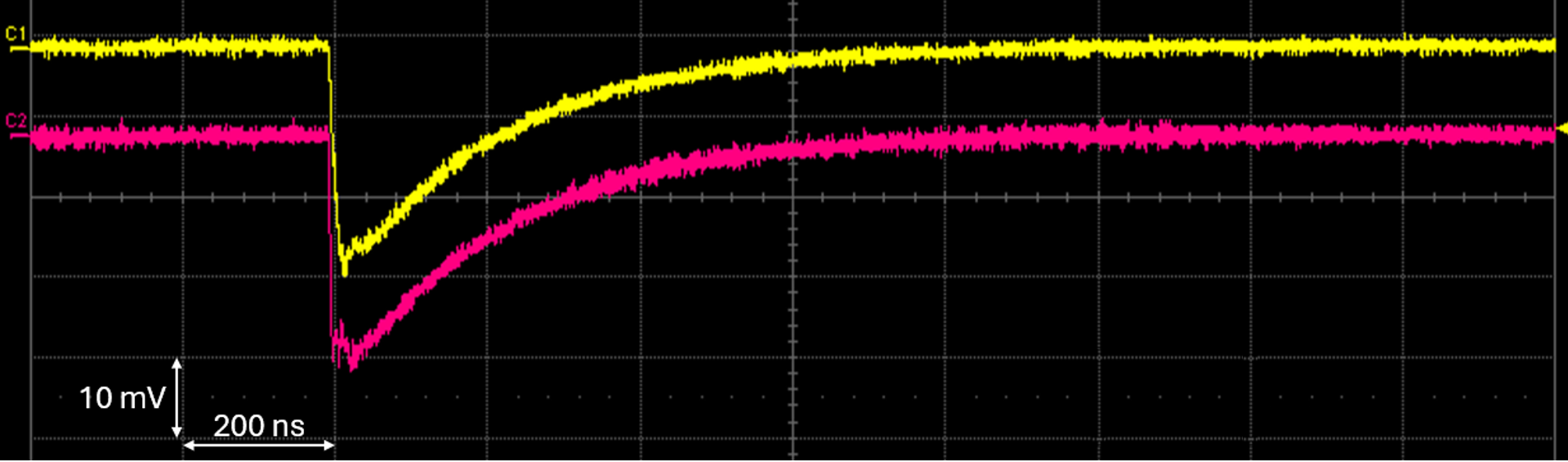}
    \caption{The average output obtained from the two channels of a scintillator when operated in parallel mode.}
    \label{output}
\end{figure}

{\red It should be noted that the current design does not fully utilize the fast response capabilities of the selected SiPMs. Additionally, due to the developed circuit architecture, in which all six SiPMs in a line are connected in parallel, their outputs do not sum as expected.
There are two possible reasons for this limitation. The first is that the activation of a single SiPM during a scintillation event induces a small voltage variation across the shared 100 \unit{\nano\farad} capacitor, leading to different gain responses among the SiPMs. The second is the fact that connecting multiple SiPMs in parallel results in a higher overall capacitance, which affects the system signal characteristics. 
Nevertheless, the high number of SiPMs considered still increases the number of scintillation events detected if compared with the 3 SiPMs PCB used for the positioning test (which has an identical circuit concept and was operated using same powering voltage and trigger on the output voltage variation). 
It is also important to note that the signal measured during the first acquisition campaign using muons as energetic particles (the one shown in Figure \ref{output}) has an excellent signal-to-noise ratio even without being amplified by the CAEN N979.}

{\red In summary, the electronics is composed by $12$x$13$=$156$ SiPMs, distributed on 13 equal PCBs, each one giving one or two independent outputs depending on the operation strategy, for a total of 13 or 26 signals. 
Since the selected fast amplifier provides 16 input channels, SARA will use one or two N979 modules, depending on the operating mode, and the required number of digitizers will be two or four. In both cases, a single CAEN DT5485PB power supply will be used thanks to the custom-made power switch.}


\subsection{Performance Optimization}

\begin{figure}[h]
    \centering
    \includegraphics[width=1\linewidth]{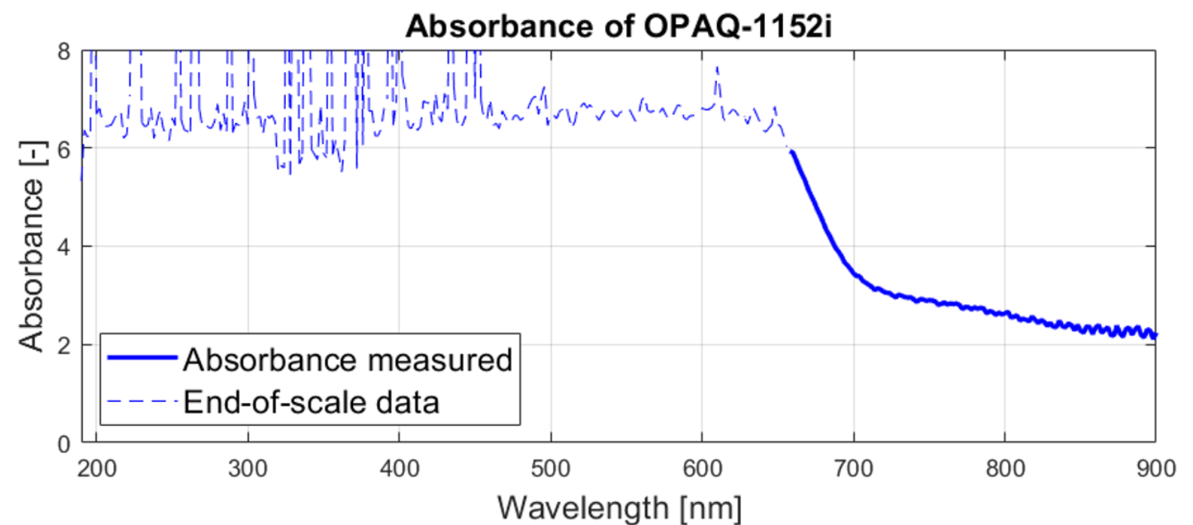}
    \caption{OPAQ-1152i absorbance {\red measured with a PerkinElmer Lambda 650 spectrophotometer}.}
    \label{absorb_emis}
\end{figure}

Regarding the performance of the scintillators, two goals were considered: to maximize the amount of light reaching the SiPMs and to prevent external light infiltrations. 
To {\red this purpose}, all the six faces of the rectangular scintillator panels {\red were} polished and wrapped with aluminum foil with the reflective side facing inward; the aluminum reflective surface {\red was} cleaned and flattened before the application. 
{\red Then}, after the integration of aluminum foil and PCB, the scintillators {\red were} covered with an obscuring material{\red, as commonly done for plastic scintillators}. Specifically, the one selected was the OPAQ-1152i from Luminis Films \cite{Opaq}, a black PVC foil designed to block visible light. 
Its selection was justified by the results {\red of its absorbance characterization.} The absorbance was {\red measured by the spectrophotometer used -- a PerkinElmer Lambda 650 -- } as:

\begin{equation}
    A(f) = log_{10}\frac{\Phi_R(f)}{\Phi_S(f)}
\end{equation}

where $\Phi_R(f)$ is the intensity as function of the frequency of a reference light ray and $\Phi_S(f)$ is the intensity of an identical light ray after it pierced the tested material. {\red The reference light was generated by the instrument combining a deuterium and an halogen lamps, whose resulting light was manipulated by a series of two monochromators to get the frequency dependence. In addition, the two identical beams were obtained by the device exploiting an internal beam splitter.}
{\red T}he results of the test {\red (}Figure \ref{absorb_emis}{\red )} {\red show} a {\red rapid increment} of the absorbance when moving toward lower wavelengths, before the instrument reaches saturation and starts to give unreliable values. 
{\red According to the} results, the chosen light protection material has an absorbance greater than 6 in the region of the visible spectrum{\red , so it is tailored for its obscuration role.}

\bigskip

\bigskip


\section{Commissioning}
\label{Commissioning}

\begin{figure}[h]
    \centering
   \includegraphics[width=0.97\linewidth]{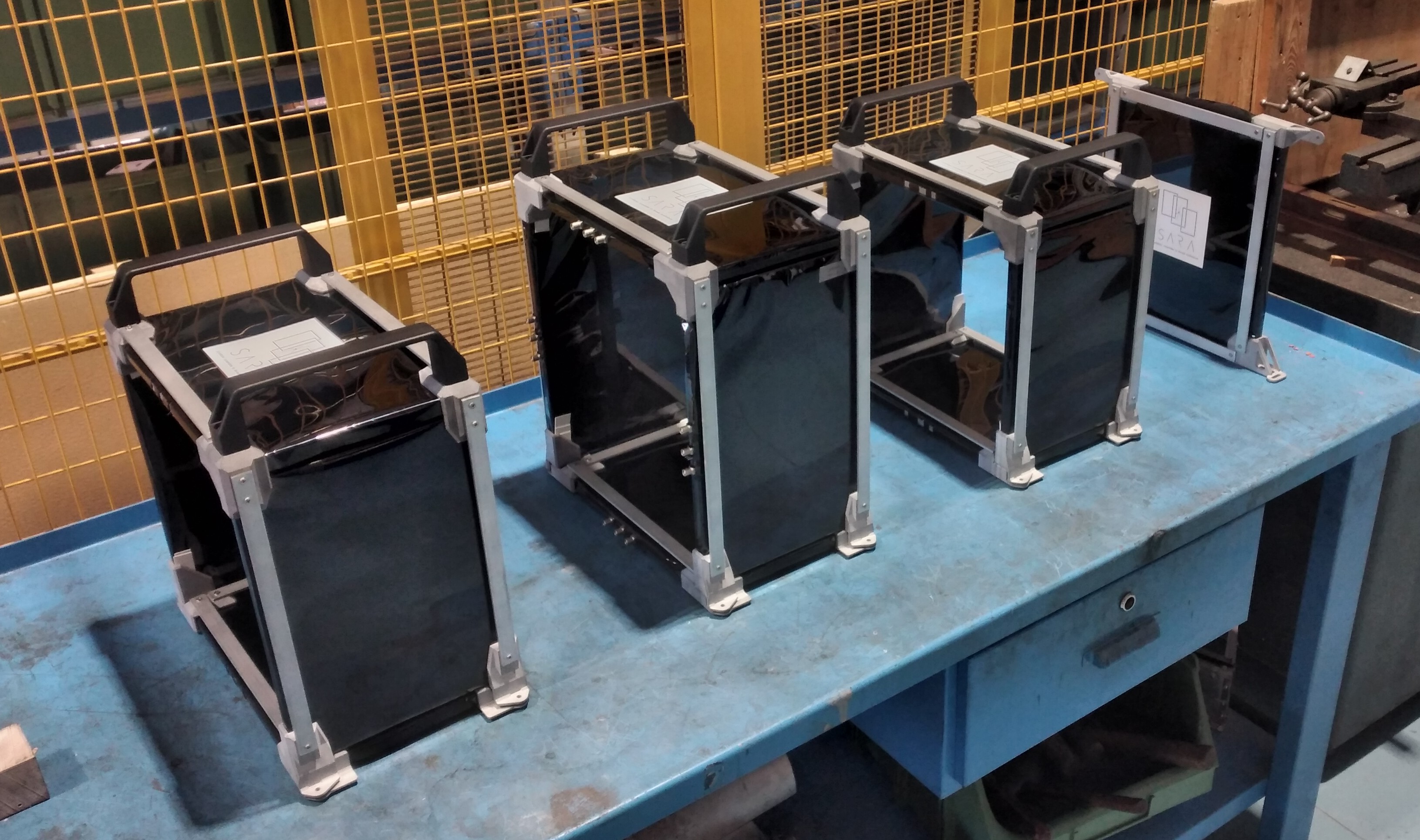}
    \caption{The instrument before the completion of the {\red PCBs} integration. }
    \label{SARA}
\end{figure}

After the conclusion of the design phase, the construction process started, yielding the results shown in the following figures. Figures \ref{commis} - 1. and 2. {\red show} some scintillators after {\red the polishing process} and some others wrapped in the aluminum foil; the rectangular holes allow the optical contact {\red between SiPMs and} scintillator. A scintillator with its detectors PCB secured with black {\red adhesive} tape and three completed scintillators {\red are shown in Figures \ref{commis} - 3. and 4.} Figures \ref{commis} - 5. and 6. show some corner elements with ensats already hot-inserted and the first completed SARA box (the central one). Finally, Figure \ref{SARA} shows the whole instrument before the completion of the {\red PCBs} integration.

\begin{figure}[h]
    \centering
    \includegraphics[width=0.89\linewidth]{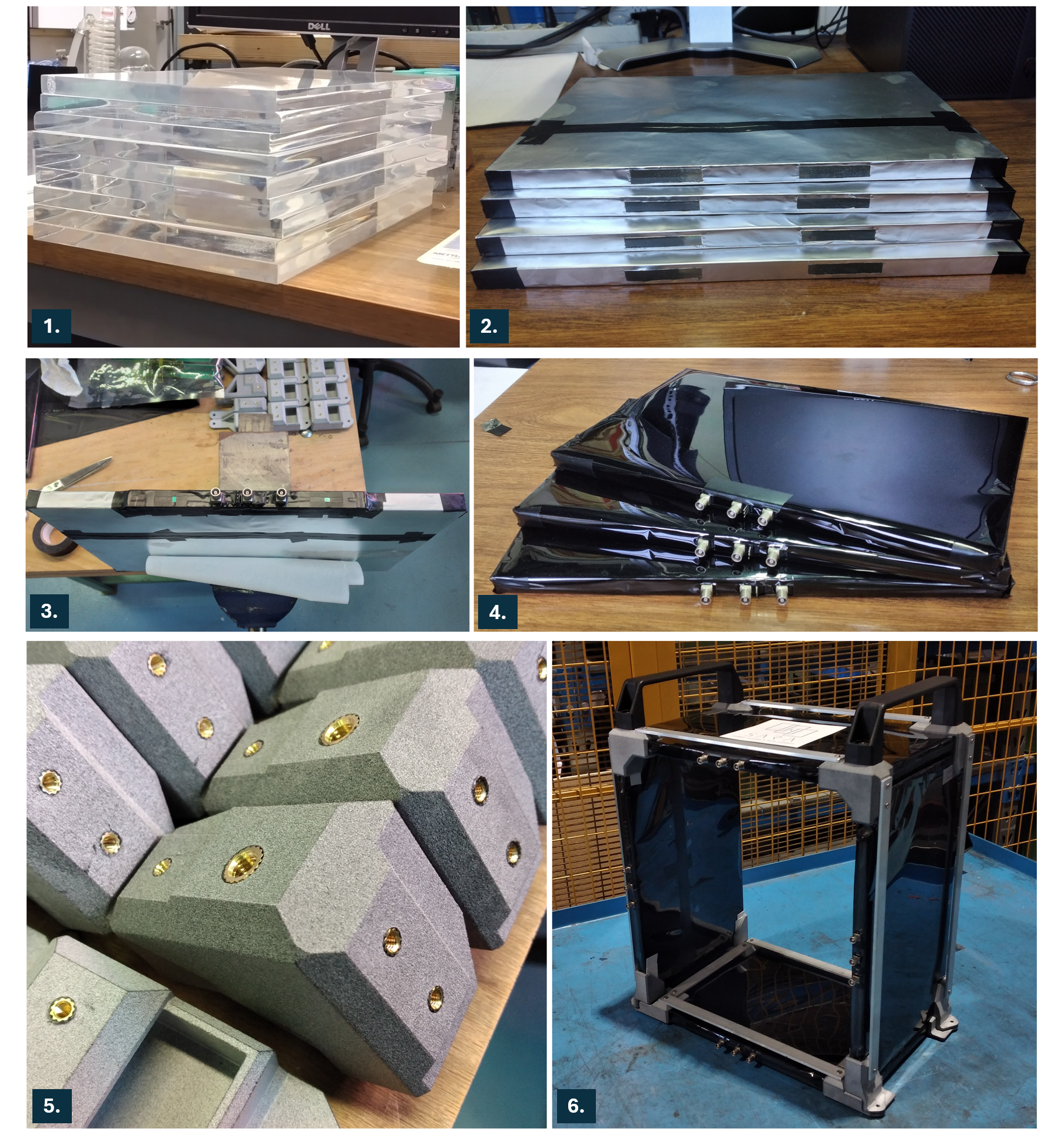}
    \caption{Construction process of the SARA scintillation detector. 1. Some of the polished scintillator panels, 2. Some scintillators wrapped in aluminum, 3. A PCB applied on a scintillator, 4. Three scintillators wrapped in OPAQ-1152i, 5. Some corner elements {\red with} ensats, 6. The first completed box.}
    \label{commis}
\end{figure}

\begin{figure}[h]
    \centering
    \includegraphics[width=1\linewidth]{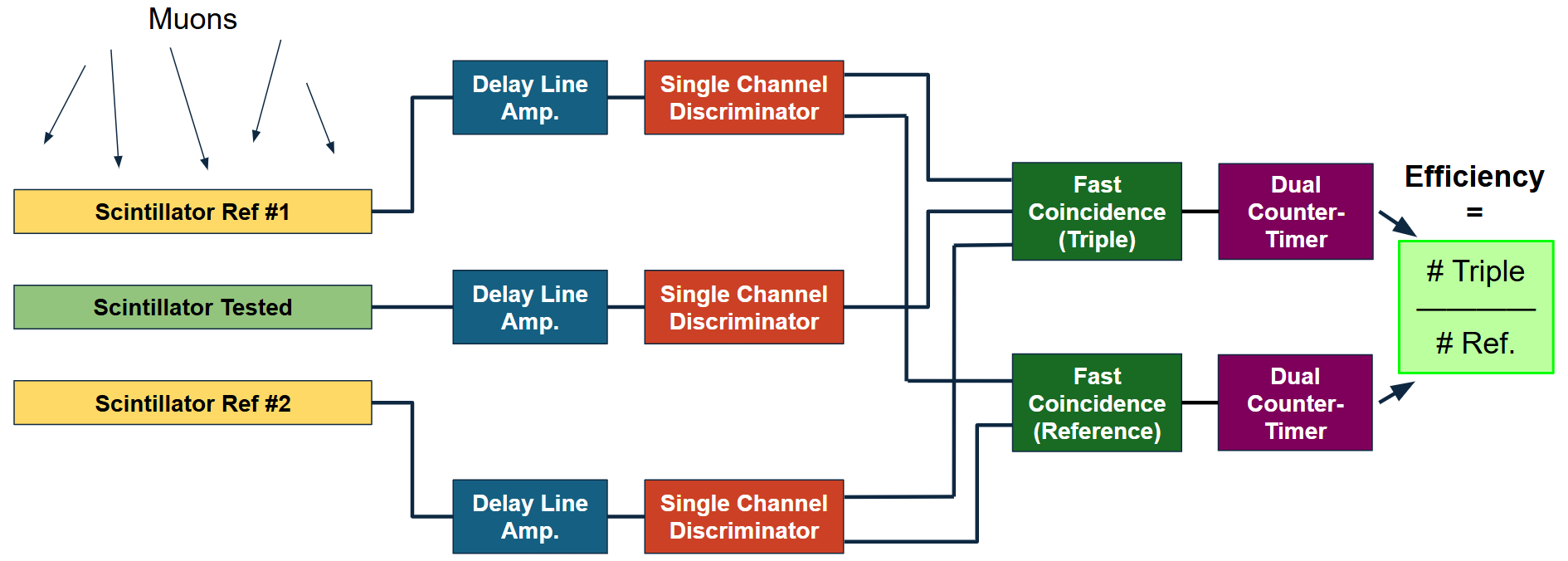}
    \caption{Scintillators efficiency measurement setup.}
    \label{eff setup}
\end{figure}

Concerning the efficiency of the scintillators, the {\red measurement procedure} exploited the fact that the scintillators composing the boxes come in just two sizes, with eight elements in the first {\red set} ($316$ \unit{\milli\meter} long panels) and four elements in the second ($270$ \unit{\milli\meter} long panels). By stacking three scintillators from the same {\red set} and using the double coincidence between the top and bottom ones as proof that a muon crossed the whole stack, the efficiency of the middle panel can be evaluated as the ratio between the number of triple coincidence{\red s} (all three scintillators triggered) and the one of double coincidences (only reference panels triggered).
This procedure was performed testing each scintillator of the two {\red sets} and the efficiency of the final panel (that comes in a unique piece) was {\red similarly measured by inserting it between two scintillators from the first set (\SI{316}{\milli\meter})}.
The coincidence counting has been achieved {\red by} assembling three identical readout channels, each one composed by a delay line amplifier (ORTEC 460 \cite{Ortec460}) and a single-channel discriminator (ORTEC 551 \cite{Ortec551}). The amplifier received the signals from a scintillator and its unipolar output was sent to the discriminator, whose lower threshold was tuned to accept only the events {\red with} highest energies in order to cut most of the noise. 
The two identical outputs of each discriminator were then sent to two ORTEC 414A Fast Coincidence \cite{Ortec414a} units, one receiving input from all three scintillators and the other {\red one} from the two reference panels {\red only}; in this way{\red ,} the two counts were referred to the same events. {\red Finally, two dual counter-timer (ORTEC 994 \cite{Ortec994}) received the signals from each coincidence unit for the counting.} The efficiency measurement setup is represented in Figure \ref{eff setup} and the results of the analysis are collected in Table \ref{efficiency}. 
{\red These values, together with those shown in Table \ref{solid} from Section \ref{scints}, were used to compute the overall efficiencies of the system modules. In particular, the latter were obtained by multiplying the average efficiency value of the scintillators of a module with the corresponding percentage of solid angle covered, obtaining the percentage of revealed annihilation products.
The results are collected in Table \ref{perc_detected}. According to them, it is possible to conclude that the system satisfies the second requirement and shows good overall efficiency for measuring coincidences, e.g., those due to the two antiparallel photons emitted during the annihilation between the $\bar H$ positron and one of the grating electrons.}

{\red The dark current was measured for} one of the PCBs, connected to the same power supply of the others, but not applied to any scintillator. The PCB was placed into a small box wrapped in multiple layers of obscuring material to prevent any possible influence of stray light. The result of the measurement, equal to \SI{1.15}{\micro\ampere}, is in agreement with the datasheet supplied by Hamamatsu.

\begin{table}[h]
    \centering
     \caption{The scintillators efficiency. {\red In the brackets is specified the position of the scintillator: two digits for the box and the acronyms Lat., Low., Upp. and Last for the position inside the module (Lateral, Lower, Upper, Last panel).}}
    \begin{tabular}{ccccc}
    \hline
        Scintillator & Efficiency measured &: &  Scintillator & Efficiency measured\\
        \hline
        1 (01\_Upp) & $0.84 \pm 0.04$ & : &  8 (02\_Low) & $0.76 \pm 0.04$ \\
        2 (02\_Lat) & $0.87 \pm 0.04$ & : &  9 (03\_Upp) & $0.77 \pm 0.04$ \\
        3 (02\_Lat) & $0.79 \pm 0.04$ & : & 10 (03\_Lat) & $0.84 \pm 0.04$ \\
        4 (01\_Low) & $0.86 \pm 0.04$ & : & 11 (03\_Lat) & $0.79 \pm 0.04$ \\
        5 (02\_Upp) & $0.82 \pm 0.04$ & : & 12 (03\_Low) & $0.81 \pm 0.04$ \\
        6 (01\_Lat) & $0.89 \pm 0.04$ & : & 13 (Last)    & $0.80 \pm 0.04$ \\
        7 (01\_Lat) & $0.84 \pm 0.04$ & : &              &                 \\
        \hline
        \end{tabular}
    \label{efficiency}
\end{table}

\begin{table}[h]
    \centering
    \caption{{\red The overall system detection capabilities. The \% of Solid Angles come from Table \ref{solid} and the averaged measured efficiencies of the modules were evaluated exploiting the values contained in Table \ref{efficiency}.}}
    \begin{tabular}{cccc}
    \hline
         Module & \% of Solid Angle & Avg. measured & \% of Detected \\
          & & Efficiency & Particles \\
         \hline
         First Box & 66.2 & 0.86 & 56.9\\
         Second Box & 64.1 & 0.81 & 51.9\\
         Third box plus Final Panel & 71.1 & 0.80 & 56.9\\
         \hline
    \end{tabular}
    
    \label{perc_detected}
\end{table}

\section{Conclusions}

{\red This article describes the design, development and commissioning of SARA (Scintillator Assemblies to Reveal Annihilations), the scintillation detector array created for the time-of-flight measurement of antihydrogen atoms within the moiré deflectometer gravity detector of the AEgIS experiment. The process aimed at realizing a simple yet effective instrument, complementary to the state-of-the-art OPHANIM detector in measuring the gravity acceleration acting on antimatter.
In more detail, the detector is composed of three independent modules, and the arrangement of the scintillators in boxes makes it possible to form coincidences between them, allowing both the detection of annihilation signals and the simultaneous minimization of noise, mainly due to cosmic muons. The two-parts box architecture effectively allows for assembly and disassembly of the boxes. Furthermore, the alignment between the various corner elements at the two parts conjunctions is satisfactory, since it guarantees the insertion of the fastening screws for all the three boxes. The choice to standardize the corner elements allows the boxes components to be symmetric, giving the possibility to rotate by $180$\unit{\degree} both the upper and lower parts with respect to the vertical symmetry axis, choosing in which direction the PCB output ports are oriented (along the deflectometer axis). This feature allows to minimize the lengths of the cables from the power supply and to the fast amplifiers. From a structural perspective, it has been demonstrated, first with the finite elements simulations performed to validate the design (described in Section \ref{FEM_sec}) and later with the inspection of the final product, that SARA is structurally stiff and able to stand all the loads it will encounter during its operational life. 
Concerning the detector performance, the signal collected during the first acquisition of muons (the one shown in Figure \ref{output}), indicates that the SiPMs coupled to the scintillators have an excellent signal-to-noise ratio, even before amplification by the fast amplifiers, and a rise time below \SI{10}{\nano\second}. 
Moving to the overall detector efficiency, its characterization showed satisfactory results. Indeed, all the three modules guarantee a percentage of detected annihilation products higher than 50\%, and the cross detection among the three modules is negligible, as can be deduced from Table \ref{solid}.  
Taking stock of everything that has been presented, it can be concluded that SARA will be able to accomplish its objective, allowing to perform the time-of-flight measurement of the antihydrogen beam, a crucial step towards the first in-beam measurement of antimatter gravitational free fall.}

\section{Acknowledgments}

The donation of plastic scintillators from the concluded ATRAP collaboration is gratefully acknowledged. {\red Professor Alberto Rotondi is here gratefully acknowledged for all his enlightening suggestions.} Special thanks are extended to Veronica Marzo for her contribution to the efficiency measurement. The project was funded by Politecnico di Milano.

\begin{flushright}
    {\small\it In memory of Sara Conte}
\end{flushright}



\end{document}